# The Powerful Model Adpredictor for Search Engine Switching Detection Challenge

Heng Gao[++ 1], Yongbao Li[++ 1], Qiudan Li[* 1], Daniel Zeng[1, 2]

[1]State Key Laboratory of Management and Control for Complex Systems

Institute of Automation, Chinese Academy of Sciences

Beijing 100190，China

[2]Department of Management Information Systems

University of Arizona, Tucson

Arizona, USA

E-mail: heng.gao@ia.ac.cn, liyongbao@fingerpass.net.cn, qiudan.li@ia.ac.cn, zeng@email.arizona.edu

## ABSTRACT

The purpose of The Switching Detection Challenge in the 2013 WSCD workshop was to predict users' search engine switching actions given records about search sessions and logs. Our solution adopted the powerful prediction model Adpredictor and utilized the method of feature engineering. We successfully applied the click through rate(CTR) prediction model Adpredictor into our solution framework, and then the discovery of effective features and the multiple classification of different switching type make our model outperforms many competitors. We achieved an AUC score of 0.84255 on the private leaderboard and ranked the 5th among all the competitors in the competition.

## Categories and Subject Descriptors

H.2.8 [**Database Management**]: Database Application – *Data mining*, H.3.3 [**Information Search and Retrieval**]: Information Search and Retrieval – *Search process.*

## 1. INTRODUCTION

With the rapid development of information technology, web users tend to acquire more information on the Internet. Therefore, search engines such as Google, Yandex and Bing become a necessity for users to satisfy their growing information acquisition demand. On the other hand, with the increasing utilization of search engines, users often face the problem of how to find the most satisfying results. Naturally, seeking for multiple search engines' results and switching within or between sessions of different search engines turn to be web users' new search habits [1]. When we observe users switching actions from one engine to another, it may imply that they require additional information and perspectives. To some extent, users' switching actions may also reflect their dissatisfaction with previous search results. For the good of search engine companies and web users, it is important for understanding of users' satisfaction with the search engine and the complexity of search.

The Switching Detection Challenge in the 2013 WSCD workshop provided us a good opportunity to study this problem. The aim of the contest was to predict users' search engine switching actions given information about the users, the queries, the URLs and the sessions. Yandex, one of the Internet giants in Russia, sponsored the task and provided one of the most valuable user s' search log datasets. Performance in the competition was measured by Area Under Curve (AUC) on a private test set, which was a subset of the whole test set. The AUC on the remainder of the test set, called the public test set, was used for the rankings on the public leaderboard.

The whole dataset consists of 10139547 unique queries, 49029185 unique URLs, 956536 unique users and 8595731 sessions. The organizers took all users that have done at least one switch within a period of 27 days and collected all search sessions for these users. This part of the data corresponds to the training period. The following 3 days were regarded as the test period and all sessions of users from the training period were test sessions. All switches from all such sessions were deleted to let participants predict their presence in these sessions. Switching actions can be monitored in three ways. First, the user might click on another search engine's URL at the result page returned in response to a navigational query seeking for another search engine. Second, Yandex provides links to major search engines at the bottom of its search engine result pages, so users can click one or several of them. Third, for the part of the users, who agreed to install Yandex browser toolbar, Yandex is able to realize that the user switched from Yandex to some another search engine at some point.

The user log represents a stream of user actions, with each line representing session metadata, a query, a click or a search engine switch, we summarize four types of the given data as follows:

1. Session metadata (*TypeOfRecord = M*): It contains 5 variables including *SessionID, Day, TypeOfRecord, USERID*, and *SwitchType*.

2. Query action (*TypeOfRecord = Q*): It contains 6 variables, including SessionID*, TimePassed, TypeOfRecord, SERPID, QueryID*, and *ListOfURLs*.

3. Click action (*TypeOfRecord = C*): It contains 5 variables, including *SessionID, TimePassed, TypeOfRecord, SERPID* , and *URLID*.

---

**++**: These authors contributed equally to this study and share the first authorship.

**\***: Corresponding author.

4. Switching action (TypeOfRecord = S): It contains 3 variables, including *SessionID, TimePassed* and *TypeOfRecord*.

The explanations of the above 10 variables in the dataset are given as follows:

- *SessionID* -The unique identifier of a search session.
- *Day* – The number of the day in the data (the entire log spans over 30 days).
- *TypeOfRecord* – It's the type of the log record. It's either a query (Q), a click (C), a switch (S), or the session metadata (M).
- *UserID* - The unique identifier of a user.
- *SwitchType* – An indicator informing if the session contains at least one switching action monitored via SERP and/or via toolbar. It's either a toolbar switch (B), a SERP switch (P), both toolbar and SERP monitored switch (H), or no monitored switches in the session (N).
- *TimePassed* – The time passed since the start of the session with the SessionID in units of time. We do not disclose how many milliseconds are in one unit of time.
- *QueryID* – The unique identifier of a query.
- *SERPID* – The unique identifier of a search engine result page at the session level (SERP).
- *URLID* – The unique identifier of an URL.
- *ListOfURLs* – The list of URLIDs ordered from left to right as they were shown to the user from the top to the bottom.

The task is to predict for each given session, if it contains a switching action anywhere in the session. The search log is supposed to be used both for training the prediction models and for prediction of the labels for the test set of sessions. The nature of this task is to make best use of the user log, and find the most important factors for users' switching actions. Although we can simplify the challenge as the classical classifying problem, it's necessary for us to make a comprehensive understanding of this problem, especially the domain knowledge in this field.

Recent years, in the domain of search engine switching actions, researchers have done much work, which helps us a lot during the feature engineering process. White et al. [2] combined large scale log-based analysis and survey data to present a study of search engine switching behavior. Their methodology characterized properties of queries, sessions, and user histories that are potentially useful in prediction. Hassan et al. [3] also investigated user behavior as a predictor of a successful search. They built novel sequence models incorporating time distributions for the task of predicting users' switching behavior, and their experiments showed the effectiveness of the sequence and time distribution models based on user behavior. Diriye et al. [4] made a deep survey on the reasons why users of search engines often abandon their searches. This paper extended previous work by studying search abandonment using both a retrospective survey and an in-situ method that captured abandonment rationales at abandonment time. It also showed that although satisfaction is a common motivator for abandonment, one-in-five abandonment instances does not relate to satisfaction. The study afforded a more complete understanding of search engine performance. There are also some researchers talking about search sequences and continuations [5, 6, 7], their work provided us a new perspective in understanding the session's sequence and its relation with users' switching actions.

Inspire by these related work, during the feature engineering process, we mainly generate 4 kinds of features, which include the user-related features, the URL-related features, the query-related features and the sequence-related features. In each of the four category features, we refine them according to the experimental results and the domain knowledge.

Adpredictor is a new Bayesian click-through rate prediction algorithm used in Micro-soft Bing search engine. As an online general Bayesian learning algorithm, it has the advantages of running fast, occupying less memory (in this application, 1GB is enough) and higher accuracy. This algorithm has showed its power in KDD CUP CTR Prediction Contest (2012) [12]. In fact, this algorithm is a special case in the TrueSkill rating system for games which could be regarded as a generalization of Elo system used in chess [9].

As we have mentioned above, we can regard the switch detection challenge as the classical binary classifying problem, in recent years, the success of the Adpredictor model called our attention. Due to the advantages of this model when compared with other classifying models, and its good appliance to the switch detection field, we adopt it as our core model in this contest.

As described in this paper, we built the Adpredictor model to solve the problem of users' switching detection. During the feature engineering process, we work around the four basic categorial features and refine each of them to select effective features. Then we utilize the Adpredictor model to validate the selected features and improve our prediction results. In the post-processing stage, we enhance our single model performance with the multi-classification idea and the rank-fusion ensemble, which includes splitting the binary classification problem into two multi-classification problems and ensemble the prediction results with the rank-fusion technology.

The remainder of this paper was organized as follows. First, we introduced our solution framework. We then described our core model Adpredictor and the selected features during the feature engineering process. Thirdly, during the experimental stage, the post-processing method and the model's performance on the test dataset was reported. In the last section, we conclude this paper.

## 2. Solution Framework

In this chapter, first of all, we propose our solution framework. Then, we tell our generated meaningful features in detail, which came out of the feature engineering process and played an important role in the precise prediction. Finally, we make efforts in the post-processing stage, which combines the task of multi-switch classification prediction and the technology of ranking fusion ensemble.

### 2.1 System Overview

The proposed system can be divided into three stages: building the validation dataset, generating meaningful features and assembling the multi-switch prediction results with the test set. In the first stage, we apply different approaches to build our own validation set, which include time-based sampling, user-based sampling, stochastic sampling and proportional sampling. The most consistent performance of the proportional sampling validation set and the test set offered us a strong suggestion to build the validation set proportionally. After building effective validation set, we generate meaningful features around the basic four categorial features. The power of the model Adpredictor not only shows in the nice prediction results, but also shows in the feature selection process. With the Adpredcitor model's high quality in selecting the effective features, in our final prediction submission result, we utilized 25 groups of refined features, which will be specified in section 2.3. In the post-processing stage, we apply ranking-fusion ensemble method to combine the results of the binary classification and the multi-classification based on users' different switching type. Figure 1 illustrates the flow of our solution framework.

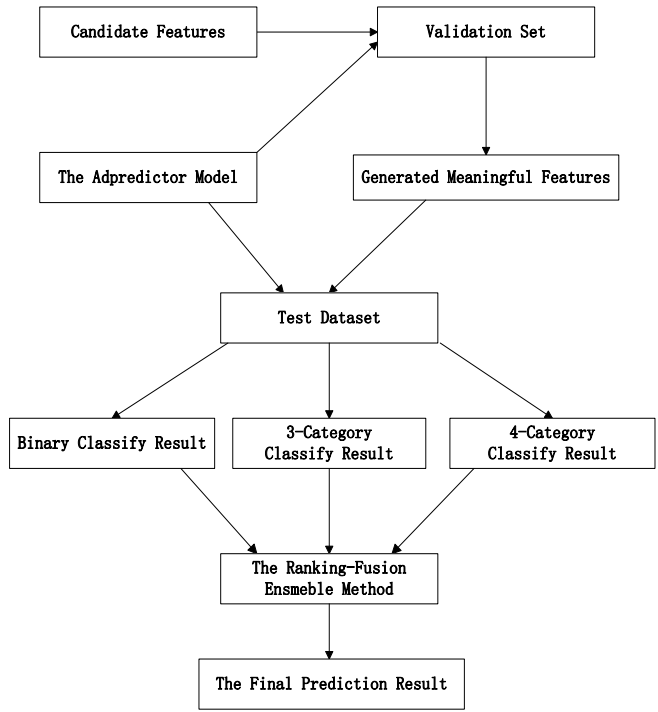

Figure 1. The framework of the switching detect prediction

## 2.2 Validation Dataset

In data mining systems, it's essential for the researchers to select proper validation dataset for the purpose of picking up nice features and models. In this challenge, users' searching behavior log of the first 27 days are regarded as the training set, and the test set comes out of the next 3 days' user log. According to the common sense, we want to utilize the later time period data in the training set for validation. However, the huge performance divergence between the validation set and the test set forces us to give up the time-based sampling method. We also tried the user-based sampling and the stochastic sampling method, and neither of them showed the consistent performance with the test set on the AUC metric. We continued trying other different methods to generate the validation set, but none of them appears more robust than the proportional sampling method. We proportionally sample 1/10 of the sessions in the training set according to the SessionID number to make the validation set. For example, we choose the SessionID number with 1, 11, 21, 31...to make the validation set. Table 1 shows the data statistics of the validation and the sub-training sets.

Table 1: Statistics of the sub-training, validation and test sets

|  | #sessions | #users | #queries | #URLs |
|---|---|---|---|---|
| Sub-training | 7071060 | 935950 | 8597114 | 42814415 |
| Validation | 785674 | 508979 | 1219540 | 8383335 |
| Test | 738997 | 254911 | 1048356 | 7271362 |

It is clear that validation set has the similar size with test set. At the same time, there are a considerable number of queries searched by different people and there are some different queries (maybe similar) with same URLs. The validation dataset plays an important role in our system framework. Firstly, we use the generated validation dataset to train the Adpredictor model and obtain the best model parameter. After that, during the process of feature engineering, we use the generated validation dataset and the trained model to find effective features among the candidate ones. We then try different kinds of feature combination methods, and the simple yet effective cumulative manner showed us the best prediction result. Finally, we take these generated features, the 3 kinds of classification form and the ranking-fusion ensemble method to aggregate the best switching detect result.

## 2.3 The Generated Features

In this section, we'd like to introduce the features we have generated in our system. In addition to the powerful model Adpredictor, successful features generation is another important factor for our team's outstanding performance in this contest. We have improved the AUC metric with more than 3 percent on the leaderboard relying on the generated features. We work around the four kinds of categorial features, and we will detail the specified features according to their different categories.

### 2.3.1 User-based features

Users play the central role in the switching detection challenge. It's important to describe users' characteristics when predicting whether they will switch search engine or not. In addition to the feature of *UserID*, which is the unique identifier of a user, we expand the user-based features with the *User_switch_ratio*, which describes users' switch ratio of different switch types in the training set.

### 2.3.2 Query-based features

The query term is another important factor deciding whether users are satisfied with the returned search results, which may further impact users' choice of switching the search engine. We collect 4 relevant features, which include *QueryID*, *Query_Count*, *Query_Duplicate*, *QueryID_Popularity*. *QueryID* is the unique identifier of a query, *Query_Count* indicates how many times a user has queried terms in the given session, *Query_Duplicate* tells us the duplicate frequency the *QueryID* shown in the given session, and we keep the *QueryID_Popularity* as the metric measuring whether a query term is a popular one.

### 2.3.3 URL-based features

The URLs appear most in a search session, which directly reflects the composition of the search results. They are also crucial to predicting users' switching actions. We gather 4 kinds of URL-based features: *URLID, Query_URLid_List, URL_Popularity* and *ClickedURL_Filtered*. *URLID* is the unique identifier of an URL, *Query_URLid_List* shows us the search results constitution of the given queries in a session, *URL_Popularity* is evaluated as the metric measuring whether a returned search URL link is a popular one, *ClickedURL_Filtered* reflects the URLs clicked by the search engine user.

### 2.3.4 Sequence-based features

As stated above, many researchers began to study how users' action sequence affected their search engine switching behavior. Users' action sequence provides us a new perspective for describing their switching behavior patterns. In our system, we come up with 5 kinds of sequence-based features: *Action_Sequence, Pattern_4gram_Normed, Pattern_5gram_Normed, Pattern_6gram_Normed* and *Pattern_7gram_Normed.* We divide the session timeline with the time interval of 100, and record users' action type of each time stamp, the whole action type sequence is regarded as the session's sequence feature, which is named as *Action_Sequence.* Take sequence "MQCQCQS" in a session as an example, Pattern_4gram_Normed takes "QCQQ" and "QCQS" as the features.

### 2.3.5 Timeline-based features

In the training and test dataset, time information connects users' various search actions, and the time information in the user log may reflect users' switching actions. For example, the longtime interval between two queries of the same session probably suggests the user's switching behavior. After he finished his first query, he may feel unsatisfied with the returned result, which promoted him to switch to another search engine, then he came back for the second time search. With the help of the model Adpredictor, we generated 4 kinds of timeline-based features, including *QueryID_Time, Query_Click_Interval, Click_Next_Query_Interval. QueryID_Time* is the time label that marks a query's start time in a session, *Query_Click_Interval* records how much time a user spends in deciding which URL to click, in contrast to it, *Click_NextQuery_Interval* offers us a clear impression of the time span between a user's last click and the next query.

### 2.3.6 Position-based features

In addition to the above categorial features, the position-based features can also characterize users' search behaviors. Imagine the picture that a user inputs a query to the search engine, and he clicks the latter URL link, we can conclude that he may not be very satisfied with the search results, in that case, he may then change to another search engine for the new search results. We generate 2 position-based features to describe users' search status, including *Cick_Position_Count* and *MRR*. We define the feature *Cick_Position_Count* as the clicked URL positions count in the same query, for those clicked URLs who positioned between 5 and 10. *MRR* is the other position-based feature we have generated, it averaged the sum of the clicked URL positions' reciprocal value.

## 3. THE MODEL ADPREDICTOR

In this section, we will give a detailed description about Adpredictor and its derivation. It is mainly used for CTR prediction for sponsored search advertising in Microsoft's Bing Search Engine [8]. From this application, we can see clearly that this model has played a good effect in switch-detect.

### 3.1 Notation

We aim to learn a mapping $\mathbf{x} \to [0, 1]$ where $\mathbf{x}$ denotes the set of feature descriptions of a user session, and the interval [0, 1] stands for the possibility of switch in a searching engine session. In this application, we use discrete multi-valued features and feature could take different values. Here, we represent the collection of features for a given sample in terms of a binary feature vector $\mathbf{x} := (\mathbf{x}_1, ..., \mathbf{x}_n)^T$,

$$\mathbf{x}_i := (x_{i,1}, ... x_{i,M_i}) \qquad \sum_{j=1}^{M_i} x_{i,j} = 1 \qquad (1)$$

each sub-vector $\mathbf{x}_i$ represents a binary 1-in-N encoding of the corresponding discrete value so that there is only one element with value 1 and the left values 0 in $\mathbf{x}_i$. We denote switch/non-switch by $y \in \{-1, +1\}$ where 1 represents switch and -1 represents non-switch.

### 3.2 Model Expression

For some parameter vector $\mathbf{w}$, a simple linear classifier classifies a point $\mathbf{x}$ according to $y = sign(\mathbf{w}^T \mathbf{x})$. Adpredictor is a generalized linear model with a probit link function as follows:

$$p(y | \mathbf{x}, \mathbf{w}) := \Phi(\frac{y \cdot \mathbf{w}^T \mathbf{x}}{\beta}) \qquad (2)$$

Here $\Phi(t) := \int_{-\infty}^{t} N(s; 0, 1) ds$ is the standardized cumulative Gaussian density probit function which results in mapping the output of the linear model in $[-\infty, +\infty]$ to [0, 1]. By using $\Phi$ instead of a step function, this likelihood tolerates small errors. The parameter $\beta$ means the steepness of the inverse link function and control the allowed "slack".

Here the weight vector $\mathbf{w}$ is supposed to obey Guassian prior distribution so that this Bayesian online learning algorithm could be solved:

$$p(\mathbf{w}) = \prod_{i=1}^{N} \prod_{j=1}^{M_i} N(w_{i,j}; \mu_{i,j}, \sigma_{i,j}^2) \qquad (3)$$

Given the sampling distributions $p(y|\mathbf{x}, \mathbf{w})$ and the prior $p(\mathbf{w})$, calculate the posterior:

$$p(\mathbf{w} | \mathbf{x}, y) = \frac{p(y | \mathbf{x}, \mathbf{w}) \cdot p(\mathbf{w})}{p(\mathbf{x} | y)}$$

$$p(\mathbf{w} | \mathbf{x}, y) \propto p(y | \mathbf{x}, \mathbf{w}) \cdot p(\mathbf{w}) \qquad (4)$$

The exact posterior over feature weights $\mathbf{w}$ could not be obtained. We could resort to factor graph model to infer the marginal belief distribution with approximate message passing. Factor graph is bipartitie graph that expresses how a "global" function of many variables factors into a product of "local" functions. We could firstly consider the joint density function $p(y, t, s, \mathbf{w} | x)$ which factorizes as:

$$p(y | t) \cdot p(t | s) \cdot p(s | \mathbf{x}, \mathbf{w}) \cdot p(\mathbf{w}) \qquad (5)$$

We can also infer the joint density function from the equation (2),

Introduce a latent variable $s$, $s := \mathbf{w}^T \mathbf{x}$ then

$$p(s) = N(s; \sum_i x_i \mu_i, \sum_i x_i^2 \sigma_i^2),$$

$$p(y = 1; \mathbf{x}, \mathbf{w}) = \Phi(\frac{\mathbf{w}^T \mathbf{x}}{\beta}) = \Phi(\frac{s}{\beta})$$

$$= \int_{-\infty}^{\frac{s}{\beta}} N(t; 0, 1) dt = \int_0^{+\infty} N(t; s, \beta^2) dt \qquad (6)$$

Introduce another variable $t$, $p(t) := N(t; s, \beta^2)$

Then

$$p(t) = N(t; \sum_i x_i \mu_i, \sum_i x_i^2 \sigma_i^2 + \beta^2)$$

$$p(y=1|\mathbf{x},\mathbf{w}) = \int_0^{+\infty} p(t)dt$$

Given the above equation, it is clear that the joint density function could be represented by the factor graph in Figure 2 and it is given by the product of all the potential functions associate with each factor. All the information about dependencies of the factors involved could be obtained from the structure of the factor graph.

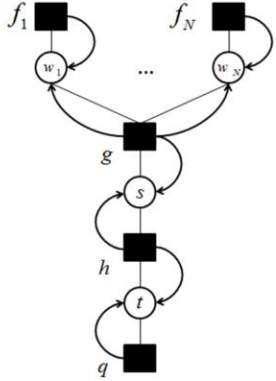

Figure 2: Factor Graph of Bayesian regression with message flow

- Factor $f_i$: sample features' weights $\mathbf{w}$ from the Gaussian prior $p(\mathbf{w})$.
- Factor $g$: introduce variable $s$ as the inner product of $\mathbf{w}^T\mathbf{x}$, so that $p(s|\mathbf{x},\mathbf{w}) := \delta(s - \mathbf{w}^T\mathbf{x})$
- Factor $h$: obtain $t$ from $s$ by adding noise subjected to zero-mean Gaussian distribution, thus $p(t|s) := N(t;s,\beta^2)$
- Factor $q$: determine $y$ by integration of density function of $t$ at zero, such that $p(y|t) := \delta(y - \text{sign}(t))$

Here the $\delta(t)$ (Dirac delta function) can be loosely thought of as a function on the real line which is zero everywhere except at the origin, where it is infinite:

$$\delta(t) = \begin{cases} +\infty, t=0 \\ 0, t \neq 0 \end{cases}, \int_{-\infty}^{+\infty} \delta(x)dx = 1$$

## 3.3 Inference

### 3.3.1 Approximation Message Passing

Factor graph allow us to transfer global function into local computations and it makes efficient computations of marginal distributions possible through the sum-product algorithm and approximate message passing. We use expectation propagation (EP) as the approximation technique so that Bayesian inference could be performed faster and more accurately. EP unifies two previous methods: assumed-density filtering and loopy belief propagation. EP approximates the belief states by only retaining expectations and, such as mean and variance, and iterates until these expectations are consistent throughout the network.

In figure 1, we have two types of marginal distributions to be computed:

- From top to bottom, given posterior $p(\mathbf{w}/\mathbf{x}, y)$ and sample feature $\mathbf{x}$, infer predictive distribution $p(y/\mathbf{x})$.
- From bottom to top, given training samples $(\mathbf{x}, y)$ and prior $p(\mathbf{w})$, infer the new posterior $p(\mathbf{w}/\mathbf{x}, y)$.

### 3.3.2 Sum-Product

According to sum-product algorithm, the results of local computations are passed as messages along the edge of factor graph followed by a simple rule. The messages computations performed could be expressed as follows:

Variable to local function:

$$m_{x \to f}(x) = \prod_{h \in n(x) \setminus \{f\}} m_{h \to x}(x) \quad (8)$$

Local function to variable:

$$m_{f \to x}(x) = (f(X_{n(f)}) \prod_{y \in n(f) \setminus \{x\}} m_{y \to f}(y)) \downarrow x \quad (9)$$

Where $n(x)$ denotes the set of factors connected to node $x$ and $m_{v \to w}(x\{v, w\})$ denotes the messages sent from node $v$ to node $w$.

When the factor graph is acyclic, the marginal $p(x)$ can be calculated from the passing messages by the following equation:

$$p(x) = \prod_{f \in n(x)} m_{f \to x}(x) \quad (10)$$

### 3.3.3 Update

State:

$$\text{Prior: } p(\mathbf{w}_{t-1}) = \prod_i N(w_{t-1}; \mu_{t-1}, \sigma^2_{t-1,j})$$

Update with new observation: $(\mathbf{x}_t, y_t)$

$$\min KL(p(\mathbf{w}_{t-1}|\mathbf{x}_t, y_t) \| p(\mathbf{w}_t))$$

$$S.T. p(\mathbf{w}_i) = \prod_i N(w_{t,i}; \mu_{t,i}, \sigma^2_{t,i})$$

We try to approximate these passing messages as well as possible by approximating the marginal $p(t_i)$ via minimization of KL divergence resulting in a Gaussian $\hat{p}(t_i)$ with the same mean and variance as $p(t_i)$.

**Step1. Calculate prior distribution**

(1) $m_{w_i \to g}: \mu_{w_i \to g} = \mu_i, \sigma^2_{w_i \to g} = \sigma^2_i$

(2) $s = \mathbf{w}^T\mathbf{x}, m_{g \to s}: \mu_{g \to s} = \sum_i x_i \mu_i, \sigma^2_{g \to s} = \sum_i x_i^2 \sigma_i^2$

(3) $m_{h \to t}: \mu_{h \to t} = \sum_i x_i \mu_i, \sigma^2_{h \to t} = \sum_i x_i^2 \sigma_i^2 + \beta^2$

For notational convenience,

$$U := \sum_i x_i \mu_i, \Sigma^2 = \sum_i x_i^2 \sigma_i^2 + \beta^2$$

$$v_t(t) := \frac{N(t;0,1)}{\Phi(t;0,1)}, \quad w(t) := v(t) \cdot [v(t) + t]$$

**Step2. Calculate $p(t)$ and $\hat{p}(t)$**

Given a known sample $(\mathbf{x}_i, y_i)$, prior distribution $m_{h \to t}(t) = N(t; U, \Sigma^2)$ According to equation (10):

$$p(t) = m_{h \to t}(t) \cdot m_{q \to t}(t), \hat{p}(t) := N(t; \mu_t, \sigma^2_t)$$

$$\text{Min } KL(p(t) \| \hat{p}(t))$$

Thus:

$$\mu_t = \frac{\int_{-\infty}^{+\infty} \delta(y_i - \text{sign}(t)) \cdot t \cdot N(t; U, \Sigma^2) dt}{\int_{-\infty}^{+\infty} \delta(y_i - \text{sign}(t)) \cdot N(t; U, \Sigma^2) dt} = y \Sigma \cdot v(\frac{y \cdot U}{\Sigma}) + U \quad (11)$$

$$\sigma_t^2 = \Sigma^2(1 - yw(\frac{U}{\Sigma})) \quad (12)$$

Here we need to use a property of Dirac delta function:

$$\int_a^b \delta(t-c) \cdot f(t)dt = f(t), a < c < b$$

**Step3. Infer** $m_{q \to t}(t)$

$m_{q \to t}(t) = p(t)/m_{h \to t}(t)$, the derivation is the same as step2.

The update equation is as follows:

$$\sigma_{q \to t}^2 = \frac{\sigma_t^2 \sigma_{h \to t}^2}{\sigma_{h \to t}^2 - \sigma_t^2} = \frac{\Sigma^2 \cdot (1 - w(\frac{y \cdot U}{\Sigma}))}{w(\frac{y \cdot U}{\Sigma})} \quad (13)$$

$$\mu_{q \to t} = \sigma_{q \to t}^2 (\frac{\mu_t}{\sigma_t^2} - \frac{\mu_{h \to t}}{\sigma_{h \to t}^2}) = \frac{y \cdot \Sigma \cdot v(\frac{y \cdot U}{\Sigma}) + Uw(\frac{y \cdot U}{\Sigma})}{w(\frac{y \cdot U}{\Sigma})} \quad (14)$$

**Step4. Infer** $m_{h \to s}$

$$m_{t \to h}(t) = m_{q \to t}(t) \quad m_{h \to s}(s) = \int h(s,t) \cdot m_{t \to h}(t)dt$$

$$\sigma_{h \to s}^2 = \frac{\Sigma^2 - (\Sigma^2 - \beta^2)w(\frac{yU}{\Sigma})}{w(\frac{yU}{\Sigma})} \quad (15)$$

$$\mu_{h \to s} = \frac{y \cdot \Sigma \cdot v(\frac{y \cdot U}{\Sigma}) + Uw(\frac{y \cdot U}{\Sigma})}{w(\frac{y \cdot U}{\Sigma})} \quad (16)$$

**Step5. Infer** $m_{g \to w_i}$

$m_{s \to g} = m_{h \to s}$, the updating equation of $w_i$ could be seen as a special case of TrueSkill rating system for games [9].

$$\sigma_{g \to w_i}^2 = \frac{\Sigma^2 - x_i^2 \sigma_i^2 w(\frac{y \cdot U}{\Sigma})}{x_i^2 \sigma_i^2 w(\frac{y \cdot U}{\Sigma})} \quad (17)$$

$$\mu_{g \to w_i} = \frac{y \cdot \Sigma \cdot v(\frac{y \cdot U}{\Sigma}) + x_i \mu_i w(\frac{y \cdot U}{\Sigma})}{x_i^2 \sigma_i^2 w(\frac{y \cdot U}{\Sigma})} \quad (18)$$

**Step6 Updating** $p(w_i)$

The update for the posterior parameter is given as below:

$p(w_i) = m_{f_i \to w_i} m_{g \to w_i}$, similar as Step2,

$$\mu_i^{new} = \mu_i + yx_i \cdot \frac{\sigma_i^2}{\Sigma} \cdot v(\frac{y \cdot \mathbf{x}^T \mathbf{\mu}}{\Sigma}) \quad (19)$$

$$\sigma_i^{2new} = \sigma_i^2 \cdot [1 - x_i \cdot \frac{\sigma_i^2}{\Sigma^2} \cdot w(\frac{y \cdot U}{\Sigma})] \quad (20)$$

## 3.4 Predictive Distribution

$$p(y \mid \mathbf{x}) = \Phi(\frac{y \cdot U}{\Sigma}) \quad (21)$$

After estimate the features' weight vector **w**, we need to infer the prediction equation. Given the $p(y|\mathbf{x},\mathbf{w})$ and $p(\mathbf{w})$, predictive distribution can be derived as the above integral.

This integral could be simplified in a closed form:

$$p(y \mid \mathbf{x}) = \int_{-\infty}^{+\infty} ... \int_{-\infty}^{+\infty} p(y \mid \mathbf{x}, \mathbf{w}) \cdot p(\mathbf{w})d\mathbf{w}$$

where

$$\mathbf{\mu} := (\mu_1, ..., \mu_N)^T, \mathbf{\sigma^2} := (\sigma_1^2, ..., \sigma_N^2)^T$$
$$U := \sum_i x_i \mu_i, \Sigma^2 = \sum_i x_i^2 \sigma_i^2 + \beta^2$$

## 4. EXPERIMENTS SETUP

In this chapter, we will detail our experiments in the switching detection challenge with three steps. First of all, we regard the switching detection task as a binary classification problem, and show the performance of the selected features on it. Then, we use the information of users' different switch type and extend the binary classification problem to the multi-classification field. In the post-processing process, we will utilize the ranking-based technology to ensemble the previous results and obtain the best prediction results.

### 4.1 Feature Selection

As stated in section 2.3, we work around the above 6 categorial features to find effective features. We select features based on the rule that the chosen features do not only improve the test set's AUC performance but also perform well in the training set. Empirically, we set the AUC improvement lower bound as 0.0005. If the AUC improvement is lower than it, we regard the added feature as the noise and discard it. Table 2 shows us the selected features set.

Table 2: The Selected Features Set

| Category | Features |
| --- | --- |
| user-based | 1.*UserID*  2. *User_switch_ratio* |
| query-based | 3.*QueryID*  4.*Query_Count*<br>5.*Query_Duplicate* 6.*QueryID_Popularity* |
| URL-based | 7.*URLID*  8.*Query_URLid_List*<br>9.*URL_Popularity*<br>10.*ClickedURL_Filtered* |
| sequence-based | 11. *Action_Sequence*<br>12-15. *Pattern_n_gram_Normed (n=4-7)* |
| timeline-based | 16. *QueryID_Time*<br>17. *Query_Click_Interval*<br>18. *Click_NextQuery_Interval* |
| position-based | 19. *Click_Position_Count*<br>20. *MRR* |

We treat the switching detection challenge as a binary classification problem, which aims at predicting whether a user switch/not switch in each session of the test set. We train the Adpredictor model with the basic features 1, 3, 7, 16 and set the parameter β as 5.0, which performs the best in the model Adpredictor with the basic features. According to the chronological order of the features' added time, we showed these selected features' AUC performance of the validation set as well as the test set in Table 3.

Table 3: The features' performance on the validation/test set

| Time Period | Added Features (detail in Table 2) | Validation-Set AUC | Test-Set. AUC |
|---|---|---|---|
| 1 | 1, 3, 7, 16 | 0.772514 | 0.799163 |
| 2 | 4,5,8,17,18 | 0.781347 | 0.809195 |
| 3 | 6,9,10 | 0.789638 | 0.815918 |
| 4 | 11,12,13,14,15 | 0.807886 | 0.823695 |
| 5 | 2,19,20 | 0.814958 | 0.829495 |

When dealing with the switching detection challenge as the switch/not switch classification problem, the Adpredictor model can obtain the best AUC value 0.829495 on the public leaderboard, which outperforms many of other competitors' prediction result. The effective refined features and the powerful Adpredictor are the most important factors. We still want to enhance our team's score, so we extend the problem with the multi-switch prediction idea in section 4.2, and ensemble the multi-switch prediction results insection 4.3, respectively.

## 4.2 Problem Extension

Notice that in this challenge, users' switch type is an indicator informing if the session contains at least one switching action monitored via SERP and/or via toolbar. The Switch Type can be refined as 4 labels, it's either a toolbar switch (B), a SERP switch (P), both toolbar and SERP monitored switch (H), or no monitored switches in the session (N). We extend the switch/not switch binary classification problem to the following 3-category classification and 4-category classification problem. As to the 3-category classification problem, we treat the blending switch type (H) in the training set as the toolbar switch (B) as well as the SERP switch (P), therefore, we have B/P/N 3-category switch type in the training set. Then we separately predict the probability users switch with the toolbar switch (B) and the SERP switch (P), and take the maximum value of them as users' switching probability. Similar to the 3-category classification problem, in the 4-category classification problem, we also treat the blending switch type (H) in the training set as the toolbar switch(B) as well as the SERP switch (P), and have the task of B/P/H/N 4-category switch prediction. At last, we take the maximum value of users' switch probability with B, P and H as the final 4-category switch type prediction result. Table 4 shows us the performance of the two type classification result.

Table 4: The AUC performance of the 3-category and 4-category classification problem

|  | Validation Set AUC | Test Set AUC |
|---|---|---|
| 3-category | 0.818739 | 0.834635 |
| 4-category | 0.821627 | 0.836106 |

We can tell from the table that the extended 3-category classification and 4-category classification idea effectively improve the AUC performance, when compared with the best AUC value of the binary classification, they raise the value by 0.004 and 0.006 separately.

## 4.3 Ranking-based Ensemble

Based on the excellent performance of the above individual results, it's reasonable to expect the further better performance of the combined prediction result. In order to leverage the results of the individual classification ones discussed in Section 4.2, we propose a ranking-based ensemble method to combine them. As we have users' switch probability in each prediction result, it is easy to combine the results by linearly assembling all the results. However, as the switch probability of different category classification may range in different intervals, some classification results may be more influential than others and even dominate the final prediction, we decide to utilize users' switch probability rank of all the test sessions to combine the yet excellent individual results.

We define rank as the ranking of a session's switch probability by ranking all of them in a descending order, thus, the session with a higher switch probability result will get a higher rank. With the adoption of the ranking method, the differences between the prediction results and the contributions of different category classifications are normalized, thus minimizing the effect of the outliers.

After we discover the sessions' switch probability rank of each classification result, we adopt the harmonic mean to ensemble the generated rankings as follows, after which we rank each session's ranking score and obtain the final submission order.

$$rank(i)\_score = \frac{3}{\frac{1}{rank_1(i)} + \frac{1}{rank_2(i)} + \frac{1}{rank_3(i)}} \quad (22)$$

Table 5 shows us the comparing results of the ensemble model and the previous binary/3-category/4-category classification model. The column of Improvement Ratio shows us how powerful the ensemble model is to enhance our challenge performance.

Table 5: The comparing results of the ensemble model and the individual models on the leaderboard

|  | Test Set AUC value | Improvement Ratio |
|---|---|---|
| binary-classification | 0.829495 | 1.6520% |
| 3-category-classification | 0.834635 | 1.0260% |
| 4-category-classification | 0.836106 | 0.8482% |
| the ensemble model | 0.843198 | — |

## 5. CONCLUSIONS

In this paper, we propose our solution to the Switching Detection Challenge of the 2013 WSCD workshop. Realizing the power of the prediction model Adpredictor, we utilize it as our core model to select effective features and predict users' search engine switch behavior. After making great progress in the binary classification area, we extend the problem to the multi-classification field, and the Adpredictor model performs even better. In the post-processing stage, we pick up the ranking-based ensemble method for the final race of the private board. With all the effective features, the powerful model Adpredictor and the reasonable ranking-based ensemble method, our team (wangzongzaimeia) achieved the AUC score of 0.842553664 on the private board, and ranked the 5th among all the contest teams.

## 6. ACKNOWLEDGMENTS

This research is supported by NSFC grants (No. 61172106, No. 71025001, No. 91024030, No. 90924302, No. 60921061, and No. 70890084), the MOH Grant (No. 2012ZX10004801), a grant from U.S. DHS Center of Excellence in Border Security and Immigration (No. 2008-ST-061-BS0002), and BJNSF(No. 4112062).